\documentclass[12pt]{iopart}

\usepackage{bm}
\usepackage{color}
\usepackage{epsfig}
\usepackage{citesort}
\usepackage{graphicx}
\usepackage{amssymb}
\eqnobysec

\newcommand{\lambdabar}{{\hbox{$\lambda$\kern-1.ex\raise+0.45ex\hbox{--}}}}

\DeclareMathAlphabet{\mathpzc}{OT1}{pzc}{m}{it}

\begin{document}

\begin{flushright}
{\large \tt
TTK-12-17}
\end{flushright}

\title{Confronting the sound speed of dark energy with future cluster surveys}

\author{Tobias Basse$^1$, Ole Eggers Bj{\ae}lde$^{1}$, Steen Hannestad$^{1}$ and Yvonne Y.~Y.~Wong$^2$}
\address{$^1$ Department of Physics and Astronomy, Aarhus University,
Ny Munkegade 120, DK-8000 Aarhus C, Denmark}
\address{$^2$ Institut f\"ur Theoretische Teilchenphysik und Kosmologie \\ RWTH Aachen,
D-52056 Aachen, Germany} \ead{\mailto{basse@phys.au.dk,
oeb@phys.au.dk, sth@phys.au.dk, yvonne.wong@physik.rwth-aachen.de}}

\begin{abstract}

Future cluster surveys will observe galaxy clusters numbering in the hundred thousands.  We consider this work how these surveys
can be used to constrain dark energy parameters: in particular, the equation of state parameter $w$ and the non-adiabatic sound speed $c_s^2$.
We demonstrate that, in combination with Cosmic Microwave Background (CMB) observations from Planck, cluster surveys such as that in the ESA Euclid project
will be able to determine a time-independent $w$  with  sub-percent precision.  Likewise, if
the dark energy sound horizon falls within the length scales probed by the cluster survey, then $c_s^2$ can be pinned down to within an order
of magnitude.  In the course of this work, we also investigate the process of dark energy virialisation in the presence of an arbitrary sound speed.
We find that dark energy clustering and virialisation can lead to dark energy contributing to the total cluster mass  at approximately the $0.1\%$ level at maximum.
\end{abstract}

\maketitle

\section{Introduction}

Current observational evidence for the accelerated expansion of the universe is compelling~\cite{Komatsu:2010fb,Kowalski:2008ez,WoodVasey:2007jb}.
This accelerated expansion is generally attributed to the presence of a ``dark energy'' component in the universe's energy budget,  a component with a strong negative pressure which dominates gravitational physics on large scales.
The nature of this dark energy is, however, still a secret the universe keeps to itself---at least for now---and much effort is being invested
to measure the properties of this dark energy in order that we may reveal its secret one day.

Because of its unknown nature, there is no shortage of explanations for what this dark energy may be~\cite{Copeland:2006wr}.  Therefore,
in order to distinguish between the various explanations, we must appeal to a range of observational tests, each probing a different aspect of dark energy dynamics. The distribution of galaxy clusters in the universe, as quantified by the cluster mass function,  is one such observational test~\cite{Sefusatti:2006eu,Takada:2007fq,Abramo:2009ne,Takada:2006xs}.

The cluster mass function (or halo mass function) is sensitive to dark energy properties primarily through their effects on the growth function.  Firstly, the dark energy equation of state $w \equiv \bar{P}_Q/\bar{\rho}_Q$, where $\bar{P}_Q$ and $\bar{\rho}_Q$ denote respectively the mean pressure and the mean energy density, affects the universal expansion rate.  This in turn alters the overall growth rate of structures, and can be probed by way of the cluster abundance as a function of redshift.
Secondly, if the dark energy is associated with a characteristic scale (such as a sound horizon),  then the growth function will also inherit this scale-dependence.  Indeed, as demonstrated in~\cite{Takada:2006xs,Basse:2010qp,Ballesteros:2010ks,Wang:2011qh},  a non-adiabatic dark energy sound speed $c_s^2$  differing from unity leads to the clustering of dark energy on scales larger than the sound horizon.  This in turn introduces a scale-dependence in the clustering of matter, which, as we shall show, ultimately translates into a distinctive mass-dependence in the cluster abundance. Note that in the case of a vanishing sound speed, dark energy is comoving with dark matter and clusters on all scales, a case that has been studied in e.g., \cite{Creminelli:2009mu,Sefusatti:2011cm}.

The purpose of this work is therefore two-fold.  Firstly, we wish to estimate as best we can the dependence of the cluster mass function on the dark energy equation of state and non-adiabatic sound speed.  To this end, we make use of the Press-Schechter formalism~\cite{Press:1973iz}, which takes as an input the linear matter power spectrum and the linear threshold density from the spherical collapse model.
Linear power spectra for nonstandard values of $w$ and $c_s^2$ are readily calculable using a Boltzmann code such as CLASS~\cite{Blas:2011rf}.  Likewise, a method for estimating the (cluster-mass dependent) linear threshold density in the case of $c_s^2 \neq 0,1$ was previously established in~\cite{Basse:2010qp}.  What remains to be determined is the fractional contribution of clustering dark energy to the total halo mass, and for this purpose we must address the issue of virialisation in the presence of clustering dark energy.
Virialisation of a combined matter and dark energy system has been studied in, e.g.,~\cite{Maor:2005hq}, in the limit $c_s^2=0$ in the context of the spherical collapse model.
In this work, we propose a simple method to track the virialisation process in a spherical collapse for an arbitrary dark energy sound speed.

Secondly, having computed the cluster abundance as a function of $w$ and $c_s^2$,  we wish to investigate the sensitivities of future cluster surveys to these parameters.
In particular,  the Euclid project  expects to observe some half a million clusters through weak gravitational lensing~\cite{Laureijs:2011mu}.  Using the appropriate survey specifications, we perform a parameter sensitivity forecast for Euclid based on the Fisher matrix approach.

The paper is organised as follows. In section~\ref{sec:II} we
review the spherical collapse model and the issue of virialisation in dark energy cosmologies. In
section~\ref{sec:III} we present the cluster mass function as functions of the dark energy parameters $w$ and $c_s^2$.   The sensitivities of Euclid to
these parameters are investigated in section~\ref{sec:IV}.  Section~\ref{sec:V} contains our conclusions.

\section{Virialisation in the spherical collapse model\label{sec:II}}

\subsection{The spherical top hat and equations of motion for the matter component}

In the spherical collapse model, a spherically symmetric overdense region is assumed to sit
on top of an otherwise uniform  background matter density field. The overdense
region has a physical radius $R_i \equiv R(\tau_i)$
at the initial (conformal) time $\tau_i$, and a uniform initial
energy density
\begin{equation}
\rho^{\rm th}_m(\tau_i)  \equiv \bar{\rho}_m(\tau_i)(1+\delta^{\rm th}_{m,i}),
\end{equation}
where $\bar{\rho}_m(\tau)$ denotes the energy density of the
background matter field.  This is our spherical ``top-hat'' perturbation, and the mass contained within the top-hat region
is given by
\begin{equation}
\label{eq:sphericalm}
M_\mathrm{halo} = \frac{4 \pi}{3} {\bar \rho}_m (\tau_i) (1 + \delta^{\rm th}_{m,i}) R_i^3
=\frac{4\pi}{3} \bar{\rho}_m(\tau_0) (1 + \delta^{\rm th}_{m,i}) X_i^3,
\end{equation}
where $\tau_0$ denotes the present time, and we have defined $X \equiv R/a$ as the comoving radius of the top hat.

The evolution of the physical top-hat radius $R$ with respect to
{\it cosmic time} $t$ is described by the familiar equation of
motion
\begin{equation}
\label{eq:ddotr}
\frac{1}{R}\frac{d^2 R}{dt^2} = - \frac{4 \pi G}{3} (\rho^{\rm th}_m + \rho^{\rm th}_Q +3 P_Q^{\rm th}),
\end{equation}
where we have incorporated in the equation a dark energy component denoted by the
subscript $Q$.  Like the matter component, the dark energy component is taken to be uniform inside the top hat
region defined by the radius $R$.  Equation~(\ref{eq:ddotr}) can be
equivalently expressed as an equation of motion for for the comoving
top-hat radius  $X$ with respect to conformal time $\tau$,
\begin{equation}
\frac{\ddot X}{X} + {\cal H} \frac{\dot X}{X} = - \frac{4 \pi G}{3}a^2 [\bar{\rho}_m \delta^{\rm th}_m  + \bar{\rho}_Q (1+3 c_s^2) \delta^{\rm th}_Q],
\label{eq:xdotdot}
\end{equation}
where $\cdot \equiv \partial/\partial \tau$,  ${\cal H}=a H$ is the
conformal Hubble parameter, and
\begin{equation}
\label{eq:cs2}
c_s^2 \equiv \frac{\delta P_Q}{\delta \rho_Q}
\end{equation}
is the square of dark energy sound speed defined in the rest frame of the dark energy
fluid. Note that in identifying $c_s$ in
equations~(\ref{eq:xdotdot}) and~(\ref{eq:cs2}) as the rest frame
sound speed, we have implicitly assumed that we are dealing only
with length scales much smaller than the Hubble length, and that the
relative velocity between the dark energy and the dark matter fluids
is much smaller than the speed of light. We also assume $c_s^2$ to
be constant in time and space.

Since the total mass of matter inside the top hat  $M_\mathrm{halo} = (4 \pi/3)
\rho_m^{\rm th} R^3$ is conserved, the top-hat matter density
contrast $\delta^{\rm th}_m$ can be easily  expressed as a
function of the top-hat radius,
\begin{eqnarray}
\label{eq:deltamth}
\delta^{\rm th}_m (\tau) & \equiv & \frac{\rho_m^{\rm th}(\tau)}{\bar{\rho}_m(\tau)}-1 \nonumber \\
&=& (1 + \delta^{\rm th}_{m,i}) \!
\left[\frac{a(\tau)}{a(\tau_i)} \frac{R_i}{R(\tau)} \right]^3  -1= (1 + \delta^{\rm th}_{m,i})  \!
\left[\frac{X_i}{X(\tau)} \right]^3  -1.
\end{eqnarray}
In the absence of dark energy perturbations, equations~(\ref{eq:xdotdot}) and~(\ref{eq:deltamth}) form a closed system for the matter component.

\subsection{Equations of motion for the dark energy component\label{sec:eom}}

On sub-horizon scales,  the pseudo-Newtonian approach~\cite{Peeblesbook} applies, so that
the (Fourier space) equations of motion for the dark energy density perturbations read~\cite{Basse:2010qp}
\begin{eqnarray}
\label{eq:linearFourier}
&&\dot{\tilde \delta}_Q + 3 {\cal H} (c_s^2 - w) \tilde\delta_Q + (1+w)
\tilde\theta_Q=0,  \\
&& \dot{\tilde\theta}_Q +(1-3 c_s^2)  {\cal H}
\tilde\theta_Q  -\frac{k^2 c_s^2}{1+w} \tilde\delta_Q+
 4 \pi G a^2 [ \bar{\rho}_m \tilde\delta_m + \bar{\rho}_Q \tilde\delta_Q]
=0,\nonumber
\end{eqnarray}
where the top-hat matter density contrast is given in Fourier space by
\begin{eqnarray}
\label{eq:Fouriermatter}
\tilde\delta_m(k,\tau) = \frac{4 \pi}{3} [(1+\delta^{\rm th}_{m,i}) X_i^3 - X^3] W(kX),
\end{eqnarray}
with the filter function
\begin{equation}
W(k X) = \frac{3}{(kX)^3} [\sin (kX) - kX \cos (kX)].
\end{equation}
The dark energy density contrast $\tilde{\delta}^{\rm lin}_Q(k,\tau)$ is then related to
 its top-hat average $\delta^{\rm th}_Q(\tau)$, as appearing in
equation~(\ref{eq:xdotdot}), via
\begin{equation}
\delta_Q^{\rm th}(\tau) = \frac{1}{2 \pi^2} \int dk \ k^2 W(kX) \tilde\delta^{\rm lin}_Q(k,\tau).
\label{eq:deltaqreal}
\end{equation}
We refer the reader to reference~\cite{Basse:2010qp} for a more detailed discussion of this formulation, but
note that  in equation~(\ref{eq:linearFourier})  we have only kept terms up to linear order in the perturbed quantities $\tilde{\delta}_Q$ and $\tilde{\theta}_Q$.
This is a reasonable approximation for $\tilde{\delta}_Q$ not exceeding unity, a condition that is generally satisfied during most of the spherical collapse process except the final moment of
collapse~\cite{Basse:2010qp}.

Equations~(\ref{eq:xdotdot}) and~(\ref{eq:linearFourier}) have been solved in~\cite{Basse:2010qp} for a range of top-hat masses to determine
the linear threshold density $\delta_c$ as a function of the halo mass.  In particular, figure~5  in the said paper shows a distinct step-like feature in $\delta_c(M)$,
with the step occurring in the vicinity of the  ``Jeans mass''
\begin{equation}
M_J\left(a\right) \equiv \frac{4\pi}{3}\bar{\rho}_m\left(a\right) \left(\frac{\lambda_J\left(a\right)}{2}\right)^3,
\label{eq:jeansmass}
\end{equation}
where $\lambda_J \equiv 2\pi/k_J$ and $k_J \equiv \mathcal{H}/c_s$ are the Jeans' length and wavenumber respectively associated with dark energy clustering.

\subsection{Dark energy contribution to the cluster mass\label{sec:cqm}}

So far we have not considered the possibility that clustering dark energy might also contribute to the mass of the collapsed object.  Following~\cite{Creminelli:2009mu}, we define
the dark energy contribution to the total mass as
\begin{equation}
\label{eq:mq}
M_Q \equiv \frac{4\pi}{3}R\left(\tau\right)^3 \delta\rho^{\mathrm{th}}_Q\left(\tau\right),
\end{equation}
with
\begin{equation}
\delta\rho^\mathrm{th}_Q\left(\tau\right) = \bar{\rho}_Q\left(\tau\right)\delta^\mathrm{th}_Q\left(\tau\right) \equiv \bar{\rho}_Q\left(\tau\right)\frac{3}{X^3}\int	_0^X{dx x^2 \delta_Q\left(x,\tau\right)},
\end{equation}
where we have smoothed the dark energy density contrast over the spherical top hat.  In defining $M_Q$ in this manner, we have implicitly assumed
that only the dark energy perturbations contribute to the total mass of the bound object, and that once gravitationally
bound, the dark energy perturbations behave  like non-relativistic matter.  It then follows that
\begin{equation}
\label{eq:masstot}
 M_\mathrm{tot} \equiv M_\mathrm{halo}+M_Q
\end{equation}
gives the total mass of the system.

\subsection{Virialisation\label{sec:vir}}

In the spherical collapse model, a gravitationally-bound halo is said to have formed in the top-hat region at the instant at which the virial theorem is first satisfied.  After this time, the
spherical region is taken to remain in a fixed configuration, with a constant radius $R_{\rm vir}$ and mass $M_{\rm vir}$ (even though formally the equations of motion dictate that the region should eventually collapse to a point of infinite density).

When only dark matter clusters and virialises, mass is conserved in the collapsing spherical region, and the virial radius $R_{\rm vir}$ can be computed relatively simply by following the kinetic and potential energies of the collapsing matter and the potential energy of the dark energy contained in the spherical region. This case of dark energy not participating in the virialisation process was examined in e.g. \cite{Maor:2005hq,Wang:2005ad}. However, when clustering dark energy also partakes in the virialisation process, the condition for virialisation changes from these simple prescriptions, because unless $c_s^2 = 0$, dark energy is not conserved in the spherical region. This necessitates a reexamination of the virialisation conditions in clustering dark energy cosmologies. Such an analysis has been carried out in~\cite{Maor:2005hq}, where both the homogeneous background as well as the clustered part of dark energy take part in  virialisation.  See also~\cite{Mota:2004pa,Horellou:2005qc,Percival:2005vm,Mota:2007zn} for other studies of dark energy virialisation.
Based on our definition of $M_Q$ in equation~(\ref{eq:mq}), however, we will assume that only the clustered part of dark energy virialises.

Our starting point is the tensor virial theorem for a system of point particles~\cite{binney},
\begin{eqnarray}\label{eq:tvt}
\frac{1}{2}\frac{d^2 I_{jk}}{dt^2} = 2K_{jk} + W_{jk},
\end{eqnarray}
where $K_{jk}$ and $W_{jk}$ are the kinetic-energy and potential-energy tensor respectively, and $I_{jk}$ is defined as
\begin{eqnarray}\label{eq:ijk}
I_{jk} \equiv \sum\limits_\alpha{m_\alpha x_{\alpha j} x_{\alpha k}},
\end{eqnarray}
with $\alpha$ running over the particles, and $\mathbf{x}_\alpha$ denoting their positions. Taking the trace of the equation~(\ref{eq:tvt})
and assuming that the system is in a steady state, we find  the well-known scalar virial theorem,
\begin{eqnarray}\label{eq:vcon}
\frac{d^2 I}{dt^2} = 0,
\end{eqnarray}
where $I \equiv \mathrm{trace}\left(\mathbf{I}\right)$. Since the trace $I$ is precisely the moment of inertia of the system,  the condition for virialisation can also be paraphrased
as the vanishing of its second time-derivative.

For a spherical system, the moment of inertia for a set of non-relativistic particles  is
\begin{equation}
I = 2M_\mathrm{tot}R^2/5,
\label{eq:inertia}
\end{equation}
 where $M_\mathrm{tot}$ denotes the mass that takes part in virialisation. Thus, in the case that mass is conserved in the spherical region (i.e., $d M_{\rm tot}/dt=0$), equation~(\ref{eq:vcon})
 together with equation~(\ref{eq:inertia}) provide a simple relation between the radius of the system and the first and second derivatives of the radius with respect to time at the time of virialisation:
\begin{eqnarray}\label{eq:vir11}
\frac{1}{R^2}\left(\frac{dR}{dt}\right)^2+\frac{1}{R}\frac{d^2R}{dt^2} = 0.
\end{eqnarray}
In an Einstein-de Sitter universe ($\Omega_m=1$) where only dark matter clusters and virialises, mass is conserved in the collapsing region, and
equation~(\ref{eq:vir11}) dictates that
virialisation happens when the radius of the system reaches half the radius at turn-around, i.e., $R_{\rm vir} = (1/2) R_{\rm ta}$,
 where turn-around refers to the instant at which the expansion of the spherical object stalls and the region starts to collapse (i.e., $dR/dt|_\mathrm{ta} = 0$).

Generalising to the case of mass non-conservation (such as when dark energy also clusters and virialises), we find from equations~(\ref{eq:vcon}) and~(\ref{eq:inertia}) the virialisation condition
\begin{eqnarray}\label{eq:vir}
\frac{1}{2M_\mathrm{tot}}\frac{d^2M_\mathrm{tot}}{dt^2}+\frac{2}{M_\mathrm{tot}R}\frac{dM_\mathrm{tot}}{dt}\frac{dR}{dt}+\frac{1}{R^2}\left(\frac{dR}{dt}\right)^2+\frac{1}{R}\frac{d^2R}{dt^2} = 0,
\end{eqnarray}
where $M_{\rm tot}$ is the total mass of the system given by equation~(\ref{eq:masstot}).  In practice,
after solving the spherical collapse, we locate the time at which equation~(\ref{eq:vir}) is satisfied. The total mass of the system at that time is then taken to be the virial mass $M_\mathrm{vir}$
of the collapsed object.

\subsection{Results\label{sec:re1}}

\begin{figure}[t]
\centering
\includegraphics[trim=0 325 220 100,clip=true,scale=1]{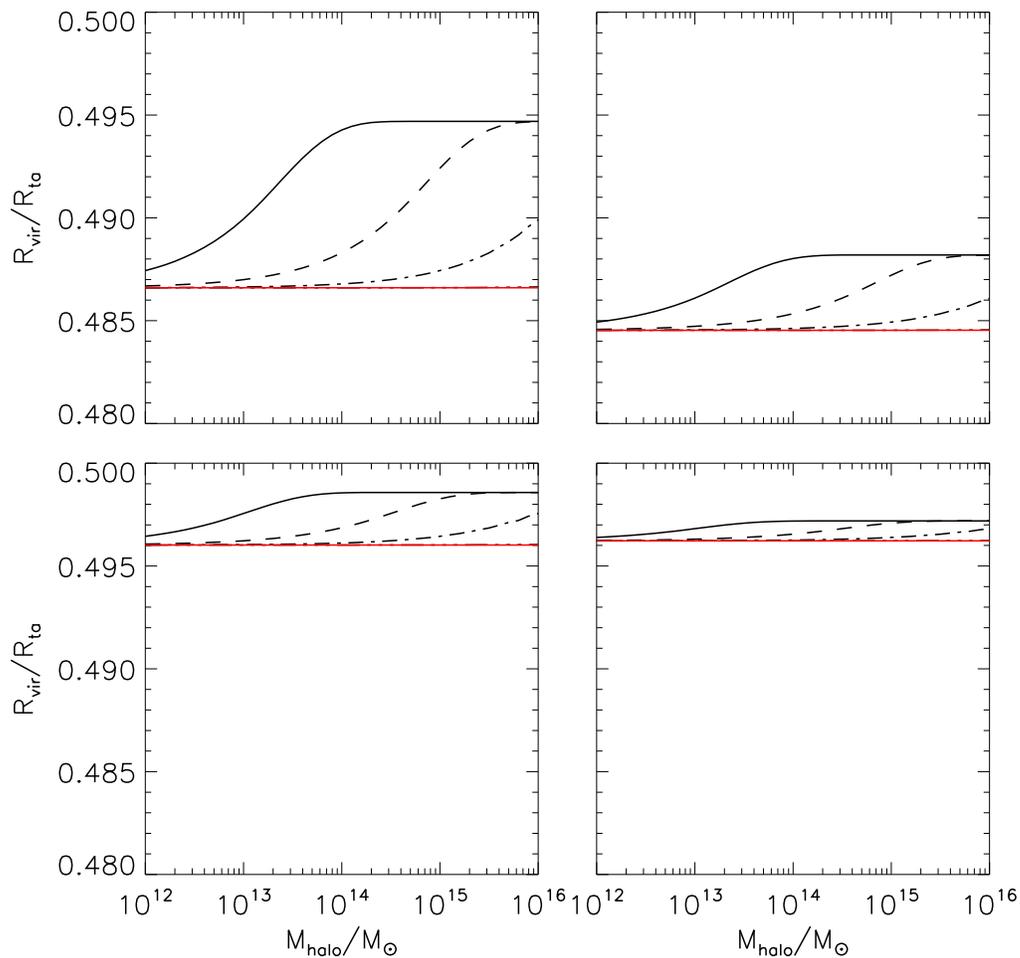}
\caption{The virial radius $R_\mathrm{vir}$, normalised to the radius at turnaround $R_\mathrm{ta}$, as a function of the dark matter halo mass $M_\mathrm{halo}$.
The left and right panels correspond to $w=-0.8$ and $w=-0.9$ respectively, while the upper and lower panels correspond to collapse at $z=0$ and at $z=1$.
 The solid black lines correspond to $c_s^2=10^{-6}$,
dashed lines to $10^{-5}$, dash-dot to $10^{-4}$, and dash-dot-dot-dot to $10^{-2}$.  The solid red line corresponds to $c_s^2=1$.}
\label{fig:RvsM}
\end{figure}

Figure~\ref{fig:RvsM} shows the radius of the spherical halo at the time of virialisation as a function of the halo's dark matter mass for several clustering dark energy cosmologies.  For clarity we have normalised the results to the radius at turn-around (i.e., the time at which $dR/dt = 0$).   As with the linear threshold density $\delta_c$ in figure~5 of \cite{Basse:2010qp}, a distinct
step-like feature at the Jeans mass can be seen in the mass-dependence of the virial radius $R_{\rm vir}$.  The case of
$c_s^2 = 10^{-6}$, for example, has $M_J\sim 10^{14} \;\mathrm{M}_\odot$, as is  evident in figure~\ref{fig:RvsM}.

The results presented in figure~\ref{fig:RvsM} can be understood in terms of a combination of several different effects: (i) the overall effect of the background homogeneous dark energy, (ii)  the evolution of the ratio of the dark energy to dark matter densities, and (iii) the effect of the dark energy clustering.

Firstly, adding a homogeneous dark energy to the universe adds a negative pressure and hence a negative contribution to the spherical system's potential energy, making it more difficult for the system to virialise.   The equation of state of dark energy tells us how strong this pressure is. Thus, the overall effect of the background dark energy is that $R_\mathrm{vir}/R_\mathrm{ta}$ becomes smaller compared to the Einstein-de Sitter case, in which $R_\mathrm{vir}/R_\mathrm{ta} = 0.5$ as discussed in the previous section.

Secondly, the more $w$ deviates from $-1$ (in the positive direction), the less rapidly  the ratio $\rho_Q/\rho_m$ evolves with time.
Fixing the ratio at $z=0$, this means that at high redshifts dark energy is present in smaller amounts in models with $w=-0.9$ than in models with $w=-0.8$.   The net effect is that
$R_\mathrm{vir}/R_\mathrm{ta}$ changes more dramatically as a function of time for $w=-0.9$ than for $w=-0.8$.  (Note that
the redshift at which dark energy starts to dominate over dark matter is $z\sim0.45$ and $z\sim0.52$ for $w=-0.9$ and $w=-0.8$ respectively.)

\begin{figure}[t]
\centering
\includegraphics[trim=0 325 220 100,clip=true,scale=1]{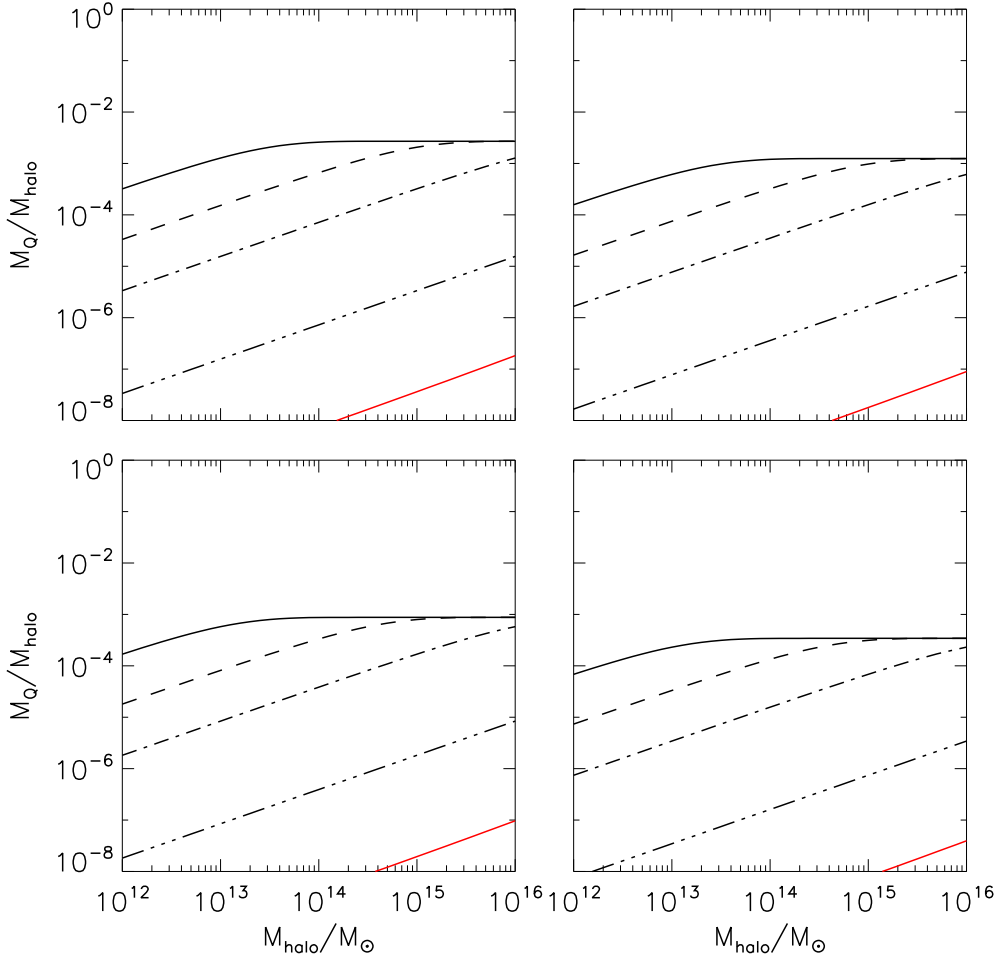}
\caption{Dark energy contribution to the total mass at virialisation relative to the dark matter mass.   The left and right panels correspond to $w=-0.8$ and $w=-0.9$ respectively, while the upper and lower panels correspond to collapse at $z=0$ and $z=1$.
 The solid black lines correspond to $c_s^2=10^{-6}$,
dashed lines to $10^{-5}$, dash-dot to $10^{-4}$, and dash-dot-dot-dot to $10^{-2}$.  The solid red line corresponds to $c_s^2=1$.}
\label{fig:MqvsM}
\end{figure}

The final effect of dark energy clustering draws $R_\mathrm{vir}/R_\mathrm{ta}$ more closely to the value for an Einstein-de Sitter universe, because a larger portion of the spherical object now participate in virialisation, and the virialisation of dark energy has been assumed in our work to proceed like non-relativistic matter.
 As previously explained, the dark energy sound speed sets the scale at which dark energy clustering becomes relevant, and the transition from weak to strong clustering is clearly evident in figure~\ref{fig:RvsM} in the case of $c_s^2 = 10^{-6}$.   The accompanying figure~\ref{fig:MqvsM} shows the contribution of dark energy to the total halo mass
 relative to the dark matter contribution at the time of virialisation.  Clearly,  dark energy clustering becomes more efficient with increasing halo mass, in the sense that the dark energy contribution to the total mass of the halo becomes increasingly important.
  However, this relative contribution does not increase indefinitely: $M_Q/M_{\rm halo}$ saturates at just above the Jeans mass, reaching a value of order $10^{-3}$.
   Comparing the $M_Q/M_{\rm halo}$ ratio at different redshifts, we see that the contribution of dark energy to the mass of the system is larger at $z=0$ than at $z=1$.  This follows simply from the fact that dark energy dominates the energy budget more at later times.

\section{The cluster mass function\label{sec:III}}

The cluster mass function is a power probe of dark energy properties.  Firstly, its redshift dependence can be used to pin down the time evolution of the linear growth function, and hence
the also the dark energy density as a function of time.  Secondly, because the cluster mass function counts the comoving number density of virialised objects per mass interval, it can
also be used to probe scale-dependent effects, such as the presence of a sound horizon due to a sound speed in the dark energy fluid.

A fully accurate cluster mass function can in principle only be computed by way of numerical simulations (e.g., $N$-body methods).  However, for the purpose of a parameter error forecast, it suffices to estimate of the generic effects of dark energy using semi-analytical methods.  To this end, we adopt the Press-Schechter formalism~\cite{Press:1973iz}, which prescribes a cluster mass function in the mass interval $[M_{\rm vir}, M_{\rm vir} +dM_{\rm vir}]$  and at redshift $z$ of the form
\begin{equation}
\frac{dn}{dM_\mathrm{vir}}(M_\mathrm{vir},z)dM_{\rm vir} = \sqrt{\frac{2}{\pi}}\frac{\bar{\rho}_{m,0}}{M_\mathrm{vir}^2}\nu \frac{d\ln \nu}{d\ln M_\mathrm{vir}}\exp{\left[-\frac{\nu^2}{2}\right]}dM_{\rm vir}.
\label{eq:excursion}
\end{equation}
Here, $\bar{\rho}_{m,0}$ is the present-day matter density, and $\nu\equiv \delta_c(z)/\sigma_m(M_{\rm vir}, z)$, with $\delta_c(z)$ the linear threshold density (for matter) at collapse and $\sigma_m(M_{\rm vir},z)$ the variance of the linear matter density field.  The latter can be computed from linear matter power spectrum  $P^\mathrm{lin}_m(k,z)$  via
\begin{equation}
 \sigma_m^2(M_\mathrm{vir},z)\equiv\frac{1}{2\pi^2}\int^\infty_0dkk^2|W(kR)|^2P^\mathrm{lin}_m(k,z),
 \label{eq:sigma}
\end{equation}
where $|W(kR)|$ is the Fourier space representation of a spherical top-hat filter of radius $R$, and
$M_\mathrm{vir}=\frac{4\pi}{3}\bar{\rho}_m(z) R^3$ relates $R$ to the virial mass of the collapsed object via the mean
matter density $\bar{\rho}_m(z)$ at redshift $z$.\footnote{On average, the dark energy does not contribute to the mass, since only the perturbations have been included in
the definition $M_Q$ in equation~(\ref{eq:mq}).}
The linear power spectrum  $P^\mathrm{lin}_m(k,z)$ can be obtained from a Boltzmann code such as CLASS~\cite{Blas:2011rf}; the publicly available version already incorporates an option for constant adiabatic dark energy sounds speeds.

Apart from its effect on the linear power spectrum, a dark energy sound speed also gives rise to a mass-dependence in the linear threshold density $\delta_c$~\cite{Basse:2010qp}.  This can be incorporated into the Press-Schechter mass function~(\ref{eq:excursion}) simply by promoting
\begin{equation}
\nu \equiv \frac{\delta_c(z)}{\sigma_m(M_{\rm vir}, z)}  \to  \frac{\delta_c(M_{\rm vir},z)}{\sigma_m(M_{\rm vir}, z)},
\end{equation}
where $\delta_c(M_{\rm vir},z)$ is computed from following a spherical collapse as discussed in section~\ref{sec:II}.  See~\cite{Basse:2010qp} for details.
Note that the definition of $M_{\rm vir}$ here includes the contribution from  clustered dark energy $M_Q$.

\subsection{Results}

\begin{figure}[t]
\centering
\includegraphics[trim=0 325 200 100,clip=true,scale=1]{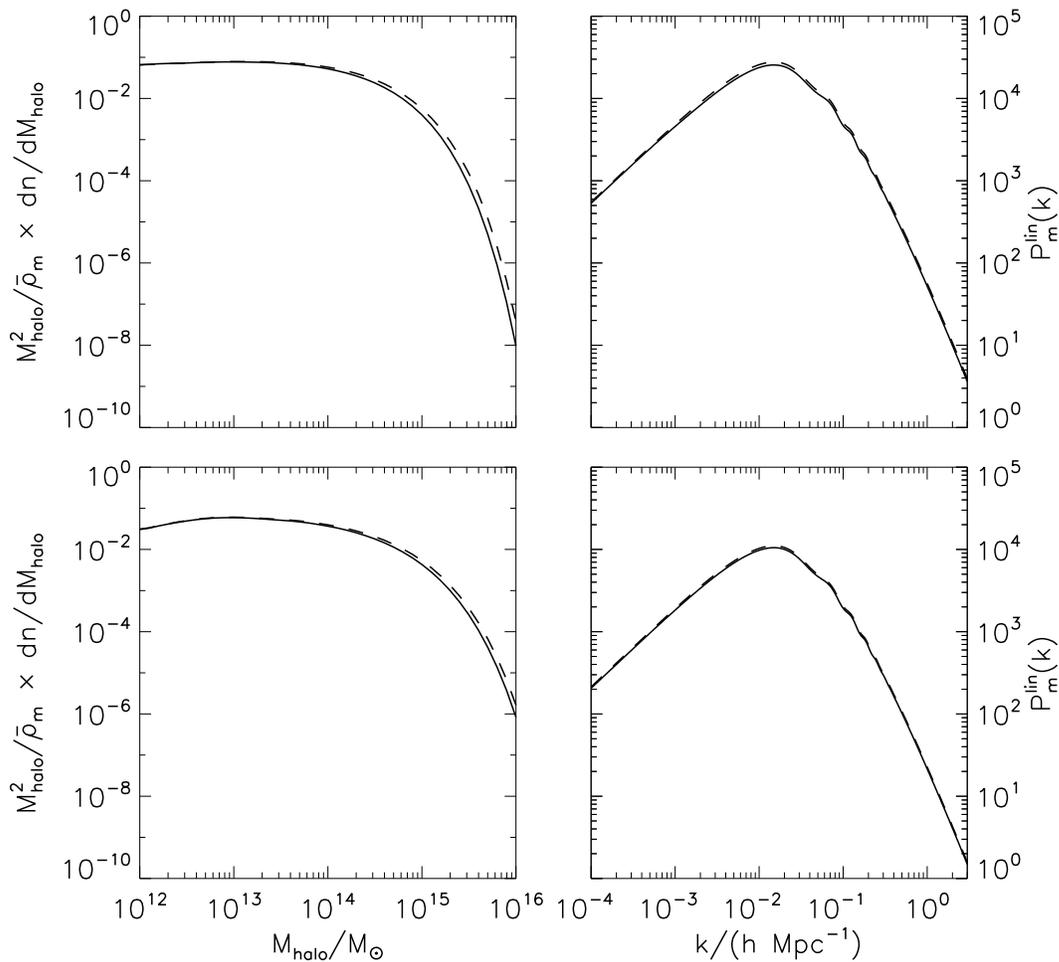}
\caption{Cluster mass functions (left) and linear matter power spectra (right) for the case of $c_s^2=1$.  Solid  lines represent a cosmology with $w=-0.8$, and
dashed lines to $w=-0.9$.  The upper and lower panels  correspond to collapse at $z = 0$ and $z = 1$ respectively.}
\label{fig:hmf_pm}
\end{figure}

Figure~\ref{fig:hmf_pm} shows the cluster mass functions at redshifts $z=0$ and $z=1$, for two cosmologies with $w=-0.9$ and $w=-0.8$ respectively, and a common sound speed $c_s^2=1$.
Other cosmological parameters have been fixed at the WMAP7 $\Lambda$CDM best-fit values~\cite{Komatsu:2010fb}.
The corresponding linear matter power spectrum are also shown in the same figure.  Dark energy dominates earlier in the case of $w=-0.8$ compared to $w=-0.9$, leading to a slightly scale-dependent suppression in the matter power spectrum of order 8\% at $z=0$ and 5\% at $z=1$.  This suppression of power due to a larger $w$ is also reflected in the cluster mass function in the form of a decrease in the number of massive objects.  At $10^{16} M_\odot$, the $w=-0.9$ cosmology has four times more objects than the $w=-0.8$ case at $z=0$, and two times at $z=1$.

\begin{figure}[t]
\centering
\includegraphics[trim=0 325 220 100,clip=true,scale=1]{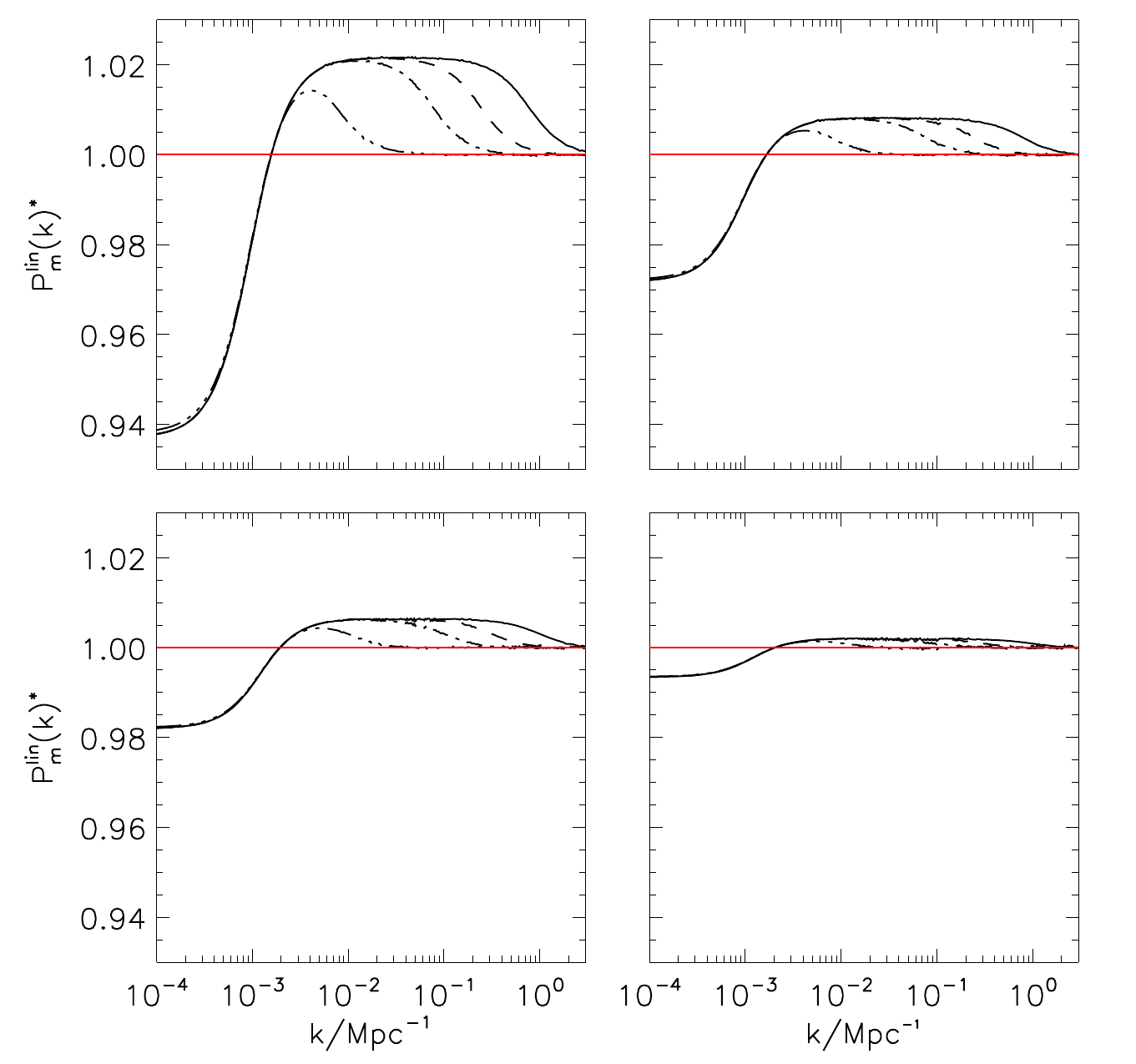}
\caption{Linear matter power spectra for various values of $c_s^2$ normalised to the $c_s^2=1$ case.  The left and right panels correspond to $w = -0.8$ and $w = -0.9$ respectively,
while the upper panels show $z=0$ and the lower ones $z=1$.  The solid black lines correspond to $c_s^2=10^{-6}$,
dashed lines to $10^{-5}$, dash-dot to $10^{-4}$, and dash-dot-dot-dot to $10^{-2}$.}
\label{fig:power}
\end{figure}

Compared with the effect of $w$, changing the dark energy sound speed from $c_s^2=1$ to a smaller value has a considerably smaller effect on both the linear matter power spectrum and the cluster mass function.  Figure~\ref{fig:power} shows the linear matter power spectra for fixed $w=-0.9$ and $w=-0.8$, while varying the dark energy sound speed between $c_s^2 = 10^{-2}$ and $10^{-6}$.  The normalised (to the $c_s^2=1$ case) spectra show that changing the dark energy sound speed induces an additional scale dependence the matter power spectrum.  In particular, on scales above the sound horizon and below the Hubble horizon, the amplitude of the matter power spectrum is enhanced.  In the $w=-0.8$ case, the enhancement is only $2 \%$
at $z=0$, and less than $1\%$ at $z=1$.   The $w=-0.9$ case sees even less enhancement.  Note that on scales close to the Hubble horizon (i.e., at very small values of $k$), a $c_s^2 \ll 1$ in fact causes the matter density contrast to decrease.  This is an artifact of the synchronous gauge used in the computation of the matter power spectrum (see, e.g.,~\cite{Ballesteros:2010ks}), and has no consequence on the observable scales.

\begin{figure}[!t]
\centering
\includegraphics[trim=0 325 220 100,clip=true,scale=1]{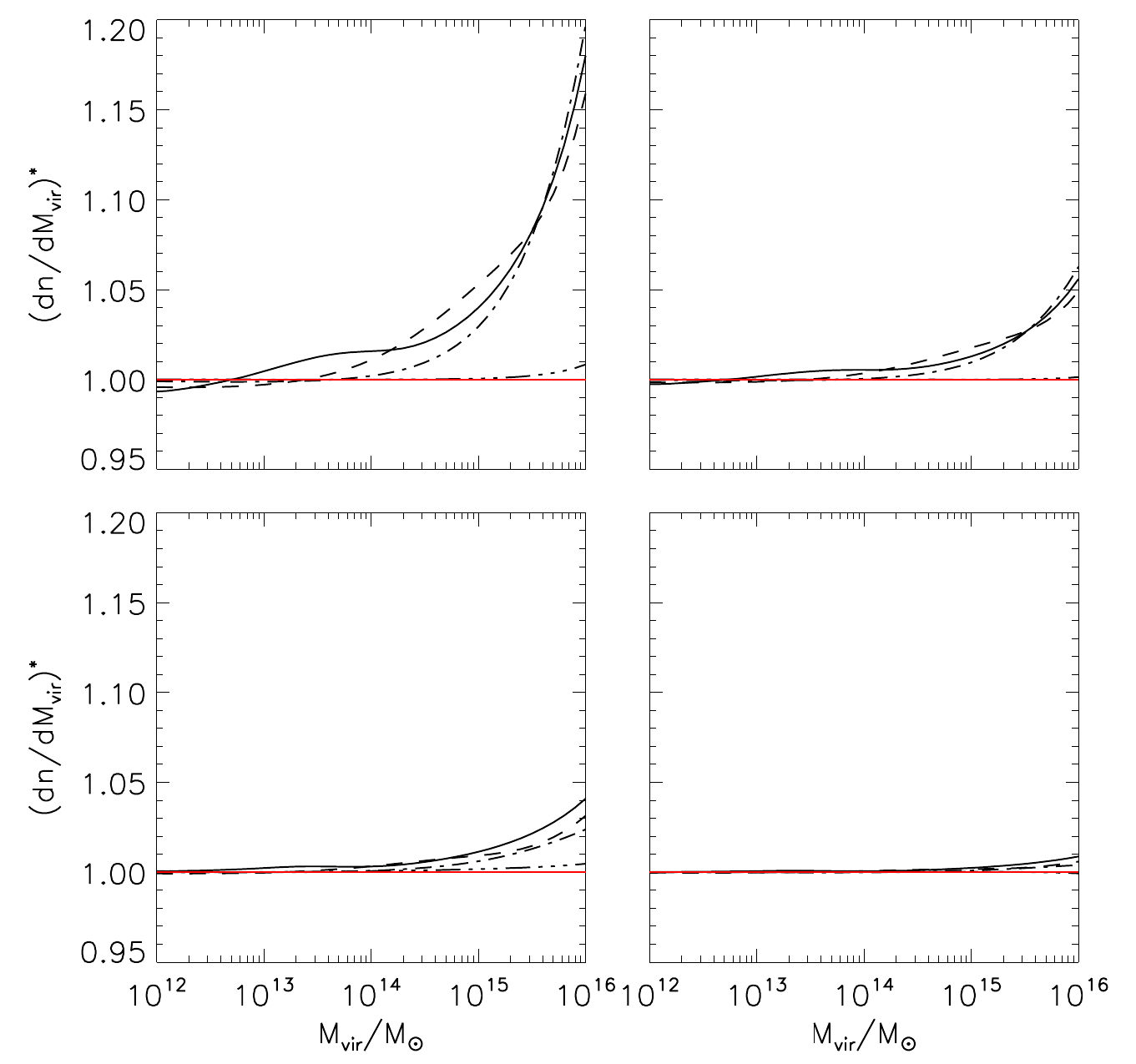}
\caption{Cluster mass functions for various values of $c_s^2$ normalised to the $c_s^2=1$ case. The left and right panels correspond to $w = -0.8$ and $w = -0.9$ respectively,
while the upper panels show $z=0$ and the lower ones $z=1$.  The solid black lines correspond to $c_s^2=10^{-6}$,
dashed lines to $10^{-5}$, dash-dot to $10^{-4}$, and dash-dot-dot-dot to $10^{-2}$.}
\label{fig:dndM}
\end{figure}

Interestingly, the effect of a dark energy sound speed appears to be somewhat more sizable on the cluster mass function.
Figure~\ref{fig:dndM} shows the cluster mass functions for a range of dark energy sound speeds,  all of which have been normalised to the $c_s^2 = 1$ case.
Evidently, clustering dark energy with sound speeds in the range $c_s^2 = 10^{-6} \to 10^{-4}$  increases the number of massive objects ($\sim 10^{16} M_\odot$) at $z=0$ by as much as $15\%$ in the $w=-0.8$ case, and $5\%$ in the $w=-0.9$ case.

The astute reader will have noticed by now that some of the cluster mass functions presented in figure~\ref{fig:dndM} have rather peculiar shapes.
This calls for an explanation. Firstly, it is clear from equation~(\ref{eq:excursion}) that the Press-Schechter mass function represents an interplay between the linear threshold density at collapse $\delta_c$ and the variance of the matter density field $\sigma_m$, with the relevant measure being the ratio $\nu=\delta_c/\sigma_m$ which generally increases with $M_{\rm vir}$.
 The shape of the cluster mass function for large masses is dominated by the exponential factor $\exp(-\nu^2/2)$ in equation~(\ref{eq:excursion}), as can be discerned from figure~\ref{fig:hmf_pm}.  Here, for a fixed value of $M_{\rm vir}$,  the role of dark energy clustering is to subtly increases $\sigma_m(M_{\rm vir})$ via its effect on the linear matter power spectrum as per equation~(\ref{eq:sigma}), especially for cluster masses  exceeding the Jeans mass $M_J$.  This leads to a lower $\nu$, and hence
 an increased number of halos on large scales.  This is the primary cause of the shapes of the cluster mass functions in figure~\ref{fig:dndM}.

\begin{figure}[t]
\centering
\includegraphics[trim=0 325 220 100,clip=true,scale=1]{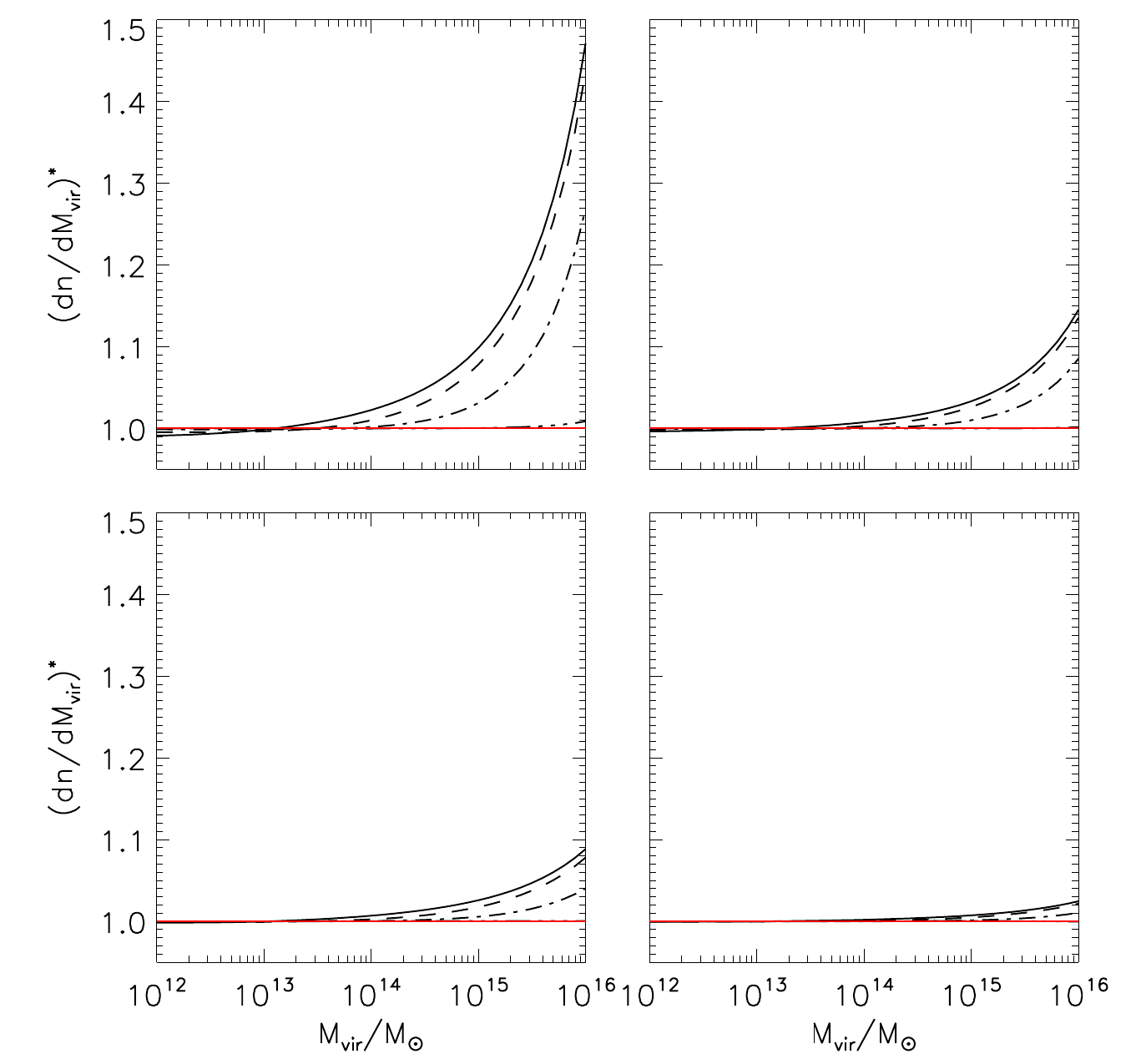}
\caption{Same as figure~\ref{fig:dndM}, but for a linear threshold density $\delta_c$ fixed at a mass-independent $\delta_c (z)= \delta_c(z,c_s^2=1)$.}
\label{fig:dndM_const}
\end{figure}

Secondly, when the dark energy sound speed deviates from $c_s^2=1$, the linear threshold density $\delta_c$ is a function of the cluster mass $M_{\rm vir}$.  This dependence on  $M_{\rm vir}$ is most pronounced at masses around the Jeans mass $M_J$; away from $M_J$,  $\delta_c$ asymptotes to $\delta_c^{\rm min}$ at $M_{\rm vir} \ll M_J$, and to $\delta_c^{\rm max}$ at $M_{\rm vir}  \gg M_J$~\cite{Basse:2010qp}.  For the case of $c_s^2=10^{-6}$ (solid line in figure~\ref{fig:dndM}), for instance, the sudden increase in $\delta_c$ occurs at around $M_J=4\times10^{14} M_{\odot}$.  This increase in $\delta_c$  compensates the growth of $\sigma_m$ with respect to $M_{\rm vir}$,
thereby causing a plateau in the cluster mass function.  At $M_{\rm vir} \gg M_J$, $\delta_c$ asymptotes to $\delta_c^{\rm max}$, and the cluster mass function recovers its exponential behaviour.

To clarify this interplay between $\delta_c(M_{\rm vir})$ and $\sigma_m(M_{\rm vir})$, we show in  figure~\ref{fig:dndM_const}  the cluster mass functions that would have arisen had the linear threshold density $\delta_c$ been fixed at $\delta_c^{\rm vir}$ (i.e., the value of $\delta_c$ at in the case of no dark energy clustering).
 Observe the absence of the aforementioned plateau.  Note also that the number of very massive clusters is a little higher here than in the figure~\ref{fig:dndM}.  This is because the higher $\delta_c$ at  large $M_{\rm vir}$ in figure~\ref{fig:dndM} leads to a stronger exponential suppression.

Finally, we note that the mass-dependence of the linear threshold density $\delta_c(M_{\rm vir})$ has been computed in this work using a linear treatment of the dark energy perturbations adopted from~\cite{Basse:2010qp}.
While this approximation works well up to the Jeans scale, it breaks down at $M > M_J$ because the dark energy density contrast inevitably exceeds unity at the time of collapse.
This break-down leads to an overestimation of the exact value of $\delta_c$ at $M > M_J$, because nonlinear dark energy clustering is expected to feed back on the matter clustering more efficiently, thereby leading to an earlier collapse for $w>-1$ than is implied by our linear treatment.
An earlier collapse in turn leads to less evolution for the linear matter density contrast,  and hence a $\delta_c$ at $M>M_J$ that is lower than that computed using our linear approximation.

Fortunately, however, the mass-dependence of $\delta_c$ works to counter the effect of dark energy clustering on $\sigma_m$, so that, as shown in figures~\ref{fig:dndM} and~\ref{fig:dndM_const}, the contrasts between the cluster mass functions for different values of $c_s^2$ are smaller when $\delta_c$ has been allowed to vary with $M_{\rm vir}$ than when it has not.
(Compare figures~\ref{fig:dndM} and~\ref{fig:dndM_const} also to  figure 7 of \cite{Creminelli:2009mu}.)  In other words, our overestimation of $\delta_c$ at $M > M_J$ leads to an {\it  underestimation} of the effect of $c_s^2$ on the cluster mass function.  Thus, when we confront our semi-analytical cluster mass functions with future observations in the next section, the sensitivities to $c_s^2$ that will be derived will be conservative.

\section{Constraining dark energy with future cluster surveys\label{sec:IV}}

In this section we perform a forecast for the sensitivity of the Euclid cluster survey to the dark energy equation of state parameter $w$ and non-adiabatic sound speed $c_s^2$
using the Fisher matrix formalism.
As we expect data of the cosmic microwave background anisotropies from Planck
 to be available long before Euclid is launched, we combine the constraints from the Euclid cluster survey with those from Planck.

\subsection{Fisher matrix formalism\label{sec:fm}}

The Fisher matrix formalism provides a quick way to compute the expected sensitivity of an experiment to cosmological parameters.
Although Monte Carlo methods are arguably a more robust forecast tool~\cite{Perotto:2006rj}, we opt not to use it here because of the time consuming real-time evaluation
of the spherical collapse.   Furthermore, the Fisher matrix has the advantage that, once computed, individual Fisher matrices for different observational probes can be combined together by simple addition.    Priors on specific parameters can likewise be incorporated into the Fisher matrix by adding $\sigma^{-2}(p_i)$ to the $ii$ entry.
We shall make use of these properties when assessing the constraining power of  Euclid and Planck, both separately and in combination.

The Fisher matrix is defined as
\begin{equation}
	F_{ij} \equiv \left\langle-\frac{\partial^2\left(\ln \mathcal{L}\right)}{\partial p_i \partial p_j}\right\rangle,
 \label{eq:fisher}
\end{equation}
where $\mathcal{L}$ is the likelihood function, and  $p_i$ is the $i$th parameter of the model.   The inverse Fisher matrix, $\left(F^{-1}\right)_{ij}$, provides the best attainable covariance matrix, regardless of the specific method used to estimate the parameters from the data~\cite{Tegmark:1996bz}. As a consequence, $\left(F^{-1}\right)_{ii}^{1/2}$ is the optimal statistical uncertainty attainable on  parameter $p_i$ after marginalisation over all other model parameters.

Assuming a Gaussian likelihood function, the Fisher matrix for the Euclid cluster survey can be constructed as~\cite{Wang:2004pk,Holder:2001db}
\begin{equation}
	F_{ij} = \sum\limits_\alpha{\frac{\partial N_\alpha}{\partial p_i}\frac{\partial N_\alpha}{\partial p_j}\frac{1}{N_\alpha}},
 \label{eq:fcc}
\end{equation}
where
\begin{equation}
	N_\alpha = \Delta\Omega \Delta z \frac{d^2 V}{d\Omega dz}\left(z_\alpha\right)\int_{M_{\min}\left(z_\alpha\right)}^{\infty}{\frac{dn\left(M,z_\alpha\right)}{dM}dM}
 \label{eq:Nalp}
\end{equation}
is the number of clusters with masses above the detection threshold $M_{\min}\left(z_\alpha\right)$ in a redshift bin centered on $z_\alpha$. In our analysis we consider 13 redshift bins from $z = 0.1$ to $z=1.4$ with bin width  $\Delta z = 0.1$, chosen so that  $\Delta z$ is twice the expected uncertainty in the determination of the cluster redshifts in the survey~\cite{Laureijs:2011mu}.
We assume the redshift bins to be top-hat functions, effectively ignoring small leakages that may arise as a result of a cluster being assigned to the wrong redshift bin.
Note that a cluster survey such as Euclid extends to $z\sim2$.   However, our analysis shows that the number of detectable clusters is negligible already at  $z\sim1.4$ since $M_{\min}$ at this redshift is of  order $10^{16} M_\odot$.
The remaining quantities in the expression~(\ref{eq:Nalp}) are
 $\Delta \Omega$  the solid angle covered by the survey, $d^2V/\left(d\Omega dz\right)(z_\alpha)$  the comoving volume element at redshift $z_\alpha$, and $dn/dM(M,z_\alpha)$ is the cluster mass function discussed in section~\ref{sec:III}.

\subsection{Euclid survey parameters\label{sec:esp}}

In order to determine the detection threshold of the weak lensing survey, i.e., $M_{\min}\left(z_\alpha\right)$ in equation~(\ref{eq:Nalp}), we adopt
 the approach of~\cite{Wang:2004pk} and~\cite{Hamana:2003ts}.
 The average shear within a Gaussian filter of angular size $\theta_G$, $\kappa_G$, is related to the mass of the cluster $M_{\rm vir}$ by~\cite{Wang:2004pk},
\begin{equation}
\kappa_G = \alpha\left(\theta_G\right)\frac{M_\mathrm{vir}/\left(\pi R_s^2\right)}{ \Sigma_\mathrm{cr}}.
 \label{eq:kappa}
\end{equation}
Assuming the mass distribution in each cluster obeys a Navarro-Frenk-White density profile with a scale radius $R_s = R_\mathrm{vir}(M_{\rm vir})/c_\mathrm{nfw}$, where $R_{\rm vir}(M_{\rm vir})$ is determined in this work via the spherical collapse,
the coefficient $\alpha(\theta_G)$ is given by
\begin{equation}
\alpha(\theta_G) = \frac{\int_0^\infty{dx\left(x/x_G^2\right)\exp\left(-x^2/x_G^2\right)f\left(x\right)}}{\ln\left(1+c_\mathrm{nfw}\right)-c_\mathrm{nfw}/\left(1+c_\mathrm{nfw}\right)}.
\label{eq:alpha}
\end{equation}
Here, $x = \theta/\theta_s$ is an angular coordinate scaled with the angular scale radius $\theta_s = R_s/d_A\left(z_l\right)$, $d_A\left(z_l\right)$  the angular diameter distance to the cluster at redshift $z_l$, $x_G \equiv \theta_G/\theta_s$  the smoothing scale, and the dimensionless surface density profile $f\left(x\right)$ is given by equation $\left(7\right)$ of~\cite{Hamana:2003ts}.

\begin{table}[t]
\caption{Euclid survey parameters.  See section~\ref{sec:esp} for an explanation.}
\centering
\vspace{2mm}
\footnotesize
\begin{tabular}{|c|rl||c|rl|}
\hline
 Parameter & Value & & Parameter & Value &\\
\hline
$\Delta\Omega$ & 20000 & deg$^2$ & $c_\mathrm{nfw}$ & 5 & \\
$\theta_G$ & 1 & arcmin & $\alpha$ & 2 &\\
$n_\mathrm{bg}$ & 30 & arcmin$^{-2}$ & $\beta$ & 2 &\\
$\sigma_\epsilon$ & 0.1 & & $z_0$ & 1 &\\
$\Delta z$ & 0.1 & & & &\\
\hline
\end{tabular}
\label{tab:Euclidpar}
\end{table}

In a geometrically flat universe, the mean inverse critical surface mass density is
\begin{equation}
\Sigma_\mathrm{cr}^{-1} = \frac{4\pi G}{c^2\left(1+z_l\right)} \chi\left(z_l\right)  \frac{\int_{z_l}^\infty{dz\, dn/dz\left(1-\chi\left(z_l\right)/\chi\left(z\right)\right)}}{{n_\mathrm{bg}}},
\label{eq:Sigma}
\end{equation}
where $\chi$ denotes the comoving radial distance to the cluster, and $dn/dz$ is the number density of source galaxies per steradian at redshift $z$, normalised such that $n_\mathrm{bg} = \int_0^\infty{dz\,dn/dz}$. We assume that  $dn/dz$ takes the form~\cite{Hannestad:2006as}
\begin{equation}
\frac{dn}{dz}dz = n_\mathrm{bg}\frac{\beta}{z_0 \Gamma\left(\frac{1+\gamma}{\beta}\right)}\left(\frac{z}{z_0}\right)^\gamma \exp\left[-\left(\frac{z}{z_0}\right)^\beta\right]dz,
\label{eq:dndz}
\end{equation}
where $\gamma$, $\beta$, and $z_0$ are chosen to fit the observed galaxy redshift distribution.  In our analysis, we fix the parameters  at $\gamma = 2$, $\beta = 2$, and $z_0 = 1$.

In order for Euclid to detect a lensing signal, the shear $\kappa_G$ must be larger than the noise.  Here we adopt a detection threshold of  $\kappa_G = 4.5\sigma_\mathrm{noise}$, where
the noise term $\sigma_{\rm noise}$ is  modeled as a ratio of the mean ellipticity dispersion of galaxies $\sigma_\epsilon$ and the number of background galaxies within the smoothing aperture $N_{\rm bg}= 4 \pi n_\mathrm{bg} \theta_G^2$~\cite{VanWaerbeke:1999wv}, i.e.,
\begin{equation}
\sigma^2_\mathrm{noise} = \frac{\sigma^2_\epsilon}{4\pi \theta_G^2 n_\mathrm{bg}}.
\label{eq:lnoise}
\end{equation}
All parameters  relevant for the Euclid analysis are listed in table~\ref{tab:Euclidpar}.  Parameter values are sourced from primarily from~\cite{Laureijs:2011mu};
parameter values missing in~\cite{Laureijs:2011mu} are taken to match those for the LSST~\cite{Wang:2004pk}.

The detection threshold $M_{\min}\left(z\right)$ at a given redshift  $z$ an now be computed from equation~(\ref{eq:kappa})
simply by finding the lowest value of $M_{\rm vir}$ at $z$ that satisfies the condition $\kappa_G > 4.5\sigma_\mathrm{noise}$.
For cosmologies in the vicinity of $\Lambda$CDM, the detection threshold is $M_{\min}\sim 10^{13} M_\odot$ at $z=0.15$, and $M_{\min}\sim 9\times 10^{15} M_\odot$ at $z=1.15$.

\subsection{Fiducial Cosmologies\label{sec:fc}}

\begin{table}[t]
\caption{Fiducial parameter values of the two models analysed in this work.}
\centering
\vspace{2mm}
\footnotesize
\begin{tabular}{|cl|cc|}
\hline
 Parameter & & Model 1 & Model 2 \\
\hline
$w$ & Dark energy equation of state & $-0.9$ & $-0.8$ \\
$\log c_s^2$ & Dark energy sound speed & $-6$ & $-6$ \\
\hline
$h$ & Hubble parameter & \multicolumn{2}{c|}{0.710} \\
$100 \Omega_b h^2$ & Physical baryon density & \multicolumn{2}{c|}{2.26} \\
$10\Omega_{cdm}$ &CDM density &  \multicolumn{2}{c|}{2.22} \\
$\log 10^{10}\Delta^2_\mathcal{R}$ & Primordial fluctuation amplitude & \multicolumn{2}{c|}{1.39} \\
$n_s$ &Scalar spectral index & \multicolumn{2}{c|}{0.963} \\
\hline
\end{tabular}
\label{tab:cosmologies}
\end{table}

We consider a 7-parameter cosmological model defined by the free parameters
\begin{equation}
\theta = \{w,\log c_s^2, h, \Omega_b h^2, \Omega_{cdm}, \log 10^{10} {\Delta_{\cal R}^2},n_s \},
\end{equation}
denoting, respectively, the dark energy equation of state, its sound speed, the Hubble parameter, the physical baryon density, the CDM density, the scalar fluctuation amplitude, and scalar spectral index.
A flat spatial geometry,  i.e., $\Omega_Q = 1 - \Omega_{cdm}-\Omega_b$, has been assumed.
We examine two different fiducial cosmologies in our analysis:
\begin{enumerate}
\item  model~1: $w =-0.9$ and $c_s^2 = 10^{-6}$,
\item  model~2: $w=-0.8$, and $c_s^2=10^{-6}$,
\end{enumerate}
where the remaining model parameters have fiducial values set by the $\Lambda$CDM best-fit of the WMAP 7-year observations~\cite{Komatsu:2010fb}.
See table~\ref{tab:cosmologies} for details.  The value of $c_s^2=10^{-6}$ has been chosen such that the characteristic sound horizon scale falls squarely within the length scales resolved by the  Euclid cluster survey (recall that the Jeans mass for $c_s^2 = 10^{-6}$ is $\sim 10^{14} M_\odot$), while $w=-0.9,-0.8$ correspond approximately to the
$1\sigma$ and $2\sigma$ lower limits inferred from the WMAP7+BAO+$H_0$ observations~\cite{Komatsu:2010fb}.

Ideally, we would also like to examine Euclid's sensitivity to dark energy models with sound speeds close to unity.  However, we find that this exercise is not feasible within the Fisher matrix framework.  This is because, as shown in figures~\ref{fig:dndM} and~\ref{fig:dndM_const}, a sound speed $c_s^2$ larger than $\sim 10^{-2}$ has essentially no observable effect on the cluster mass function on the mass scales probed by the Euclid cluster survey ($10^{13} M_\odot < M < 10^{16} M_\odot$).  The corresponding likelihood function is virtually flat in the $c_s^2$ direction in the neighbourhood of the fiducial model, which is ultimately translated by the Fisher matrix  into unbounded or unrealistically lax constraints on $c_s^2$
(recall that a Fisher matrix forecast is based upon evaluating the curvature of the likelihood function locally at the fiducial model).  The failure of the formalism to take into account the full extent of the likelihood function is a severe drawback of the Fisher matrix approach,
\footnote{Although the Fisher matrix approach is local in nature, it does return reasonable estimates if the likelihood function does not deviate too much from a (multivariate) Gaussian distribution, for which the curvature in a chosen direction is constant no matter the distance from the best-fit.}
which can  be remedied only with a full Monte Carlo simulation~\cite{Perotto:2006rj}.  We defer this exercise to a later study.

\subsection{Constraints from Euclid alone}

Table~\ref{tab:resultsEuclid} shows the $1 \sigma$ sensitivities to various cosmological parameters attainable by the Euclid cluster survey alone.   Note that when deriving these numbers, we have included the priors $\sigma\left(\Omega_b h^2\right) = 0.001$ and $\sigma\left(n_s\right) = 0.02$ on $\Omega_b h^2$ and $n_s$ respectively,
because the Euclid cluster survey  has no particular constraining power for these parameters.  These prior ranges correspond approximately to twice the $1 \sigma$ uncertainties inferred from the WMAP 7-year data~\cite{Komatsu:2010fb}.

\begin{table}[t]
\caption{$1\sigma$ sensitivities of the Euclid cluster survey to various cosmological parameters in the two fiducial models considered in this work.
  Note that the
uncertainties in $100\Omega_b h^2$ and $n_s$ are identically the priors we set on these parameters using the WMAP 7-year data.}
\centering
\footnotesize
\vspace{2mm}
\begin{tabular}{|c|cc|}
\hline
 Parameter & Model 1 & Model 2 \\
\hline
$w$  & 0.054 & 0.065 \\
$\log c_s^2$ & 1.19 & 0.46 \\
$h$  & 0.074 & 0.089 \\
$100\Omega_b h^2$ & 0.10 & 0.10 \\
$10\Omega_{cdm}$  & 0.039 & 0.038 \\
$\log 10^{10}\Delta^2_\mathcal{R}$ & 0.12 & 0.14 \\
$n_s$ & 0.020 & 0.020 \\
\hline
\end{tabular}
\label{tab:resultsEuclid}
\end{table}

The Euclid cluster survey alone is already fairly sensitive to the dark energy equation of state and hence the evolution of the dark energy density,
with a $1 \sigma$ sensitivity to $w$ of better than $10\%$ in both models~1 and~2.  These sensitivities should be compared with the uncertainties on $w$ from the
current generation of measurements:  for a flat model with $c_s^2=1$, a constant $w$ parameter can be constrained to the same level of uncertainty ($\sigma(w) \sim 0.05$) only with
a combination of the WMAP 7-year data, measurements of the baryon acoustic oscillations, and type Ia supernovae; dropping the supernova measurements immediately enlarges the uncertainty by at least a factor of two~\cite{Komatsu:2010fb}.

Our results also show that, if the fiducial dark energy sound speed falls well within the observable range of the cluster survey, then the model can be distinguished from $c_s^2=0,1$
with a good degree of certainty.  In our example of  $\log c_s^2 = -6$, the $1 \sigma$ sensitivity of Euclid to $\log c_s^2$ is $1.19$ and $0.46$ for fiducial $w$ values of $-0.9$ and $-0.8$ respectively: the sensitivity to $\log c_s^2$ is better in the latter model because dark energy clustering is generally suppressed by a factor $\left(1+w\right)$ relative to dark matter clustering.

\subsection{Constraints from Planck+Euclid}

\begin{table}[t]
\caption{Survey specifications of Planck in three frequency channels~\cite{Rocha:2003gc}.}
\vspace{2mm}
\centering
\footnotesize
\begin{tabular}{|c|ccc|}
 \hline
 & 100~GHz & 143~GHz & 217~GHz \\
 \hline
$\theta_c$ (arcmin) & 10.7 & 8.0 & 5.5 \\
$\sigma_{\mathrm{T},c}$ ($\mu\mathrm{K}$) & 5.4 & 6.0 & 13.1 \\
$\sigma_{\mathrm{E},c}$ ($\mu\mathrm{K}$) & -- & 11.4 & 26.7 \\
$\ell_c$ & 757 & 1012 & 1472 \\
\hline
\end{tabular}
\label{tab:planck}
\end{table}

Planck measures the temperature and polarisation anisotropies of the Cosmic Microwave Background (CMB).   These measurements are quantified at the simplest level in terms of
 auto- and cross-correlations of the temperature $T$ and  $E$-/$B$-type polarisation fluctuations.
 Neglecting $B$-mode polarisation, the CMB Fisher matrix is given by~\cite{Zaldarriaga:1996xe}
\begin{equation}
F_{ij} = \sum\limits_\ell{\sum\limits_{X,Y}{\frac{\partial C_{X,\ell}}{\partial p_i}\mathrm{Cov}^{-1}\left(C_{X,\ell},C_{Y,\ell}\right)\frac{\partial C_{Y,\ell}}{\partial p_j}}},
\label{eq:cmbfisher}
\end{equation}
where $X$ and $Y$ runs over $TT$, $EE$ and $TE$, and the entries in the symmetric covariance matrix are given by
\begin{eqnarray}
\nonumber\mathrm{Cov}\left(C_{\mathrm{TT},\ell},C_{\mathrm{TT},\ell}\right) = \frac{2}{\left(2\ell+1\right)f_\mathrm{sky}}\left(C_{\mathrm{TT},\ell}+B_{\mathrm{T},\ell}^{-2}\right)^2, \\
\nonumber\mathrm{Cov}\left(C_{\mathrm{EE},\ell},C_{\mathrm{EE},\ell}\right) = \frac{2}{\left(2\ell+1\right)f_\mathrm{sky}}\left(C_{\mathrm{EE},\ell}+B_{\mathrm{E},\ell}^{-2}\right)^2, \\
\nonumber\mathrm{Cov}\left(C_{\mathrm{TE},\ell},C_{\mathrm{TE},\ell}\right) = \frac{1}{\left(2\ell+1\right)f_\mathrm{sky}} \! \left[C_{\mathrm{TE},\ell}^2+\left(C_{\mathrm{TT},\ell}+B_{\mathrm{T},\ell}^{-2}\right) \! \left(C_{\mathrm{EE},\ell}+B_{\mathrm{E},\ell}^{-2}\right)\right],\\
\nonumber\mathrm{Cov}\left(C_{\mathrm{TT},\ell},C_{\mathrm{EE},\ell}\right) = \frac{2}{\left(2\ell+1\right)f_\mathrm{sky}}C_{\mathrm{TE},\ell}^2,\\
\nonumber\mathrm{Cov}\left(C_{\mathrm{TT},\ell},C_{\mathrm{TE},\ell}\right) = \frac{2}{\left(2\ell+1\right)f_\mathrm{sky}}C_{\mathrm{TE},\ell}\left(C_{\mathrm{TT},\ell}+B_{\mathrm{T},\ell}^{-2}\right),\\
\mathrm{Cov}\left(C_{\mathrm{EE},\ell},C_{\mathrm{TE},\ell}\right) = \frac{2}{\left(2\ell+1\right)f_\mathrm{sky}}C_{\mathrm{TE},\ell}\left(C_{\mathrm{EE},\ell}+B_{\mathrm{E},\ell}^{-2}\right).
\label{eq:cmbcov}
\end{eqnarray}
Here, $f_\mathrm{sky}$ is the fraction of the sky remaining after removal of foregrounds and galactic plane contaminations, etc., and $B_{X,\ell}$ denotes the expected error of the experimental apparatus~\cite{Efstathiou:1998xx,Knox:1995dq},
\begin{equation}
B_{X,\ell}^2=\sum\limits_c{\left(\sigma_{X,c}\theta_{c}\right)^{-2} e^{-\ell\left(\ell+1\right)/\ell_c^2}},
\label{eq:cmberror}
\end{equation}
where the index $c$  labels the different frequency channels, $\sigma_{X,c}^2$  is the variance of the instrumental noise in the temperature/polarisation measurement,
$\theta_{c}$ is the width of the beam assuming a Gaussian profile, and $\ell_c \equiv 2\sqrt{2\ln2}/\theta_c$ is the corresponding cut-off multipole.   Table~\ref{tab:planck} lists the values of these quantities specific to Planck's observations.

\begin{table}[t]
\centering
\caption{$1\sigma$ sensitivities to various cosmological parameters from the Planck temperature and polarisation measurements ($TT$, $EE$ and $TE$). }
\vspace{2mm}
\footnotesize
\begin{tabular}{|c|cc|}
\hline
Parameter & Model 1 & Model 2 \\
\hline
$\sigma\left(w\right)$ & 0.21 & 0.21 \\
$\sigma\left(\log c_s^2 \right)$ & 5340.77 & 948.04 \\
$\sigma\left(h\right)$ & 0.082 & 0.090 \\
$\sigma\left(100\Omega_b h^2\right)$ & 0.026 & 0.025 \\
$\sigma\left(10\Omega_{cdm}\right)$ & 0.52 & 0.57 \\
$\sigma\left(\log 10^{10}\Delta^2_\mathcal{R}\right)$ & 0.010 & 0.010 \\
$\sigma\left(n_s\right)$ & 0.0067 & 0.0063 \\
\hline
\end{tabular}
\label{tab:planck_results}
\end{table}

Table~\ref{tab:planck_results} shows the $1 \sigma$ sensitivities of Planck to the parameters of our two fiducial models.  Compared with the Euclid-only results in table~\ref{tab:resultsEuclid}, it is immediately clear that Planck has very little sensitivity to dark energy parameters compared with cluster surveys; the $1 \sigma$ sensitivity to $w$ is only about $20\%$, while $\log c_s^2$ is completely unconstrained, consistent with previous expectations (e.g.,~\cite{DeDeo:2003te,Bean:2003fb,Hannestad:2005ak}).  Planck's sensitivity to $h$ is comparable to that of the Euclid cluster survey ($\sim 10\%$), but the latter performs ten times better for the $\Omega_{cdm}$ measurement.

\begin{figure}[t]
\centering
\includegraphics[trim=0 370 200 100,clip=true,scale=0.7]{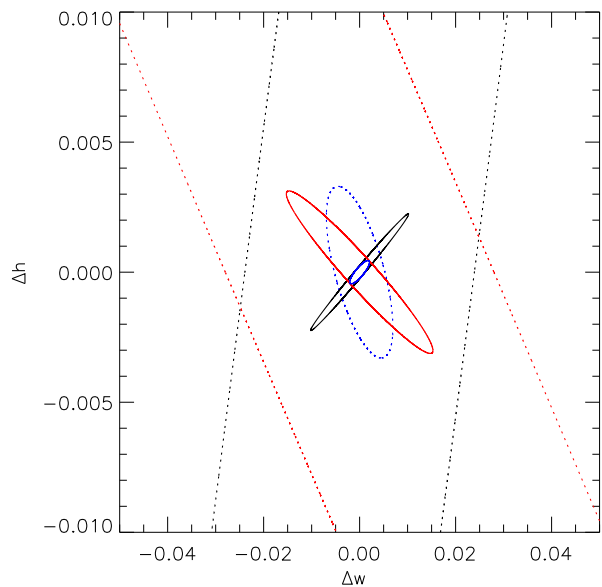}
\caption{$1 \sigma$ sensitivities in the  $\left(h,w\right)$-parameter space for the Euclid cluster survey (black lines), Planck (red), and combined Planck+Euclid (blue),
assuming fiducial model~2.
The dotted lines represent sensitivities after marginalising over all other cosmological parameters, while the solid lines correspond to no marginalisation (i.e., parameters
not shown in this plot are kept fixed at their fiducial values).}
\label{fig:degen}
\end{figure}

What is more interesting is when observations from Planck and the Euclid cluster survey are used in combination.
The combined $1 \sigma$ sensitivities of Planck and the Euclid cluster survey are displayed in table~\ref{tab:Euclid+planck}.  Compared with the expectations from Planck or
Euclid alone, we see that  Planck+Euclid improves the sensitivities to some parameters, especially $h$ and $w$, by more than tenfold; the $1 \sigma$ sensitivities to $w$ and $h$ from the combined analysis are both now better than $1 \%$.  Such a significant improvement shows that Planck and the Euclid cluster survey are individually sensitive to different degenerate combinations of these parameters.  However, once these observations are used in combination, the degeneracies are completely broken.  Figure~\ref{fig:degen} illustrates how the degeneracies in $h$ and $w$ inherent  in the Planck and in the Euclid observations
can be broken by a combined analysis.

\begin{table}[t]
\caption{$1\sigma$ sensitivities to various cosmological parameters from combining the Euclid cluster survey and the Planck temperature and polarisation measurements ($TT$, $EE$ and $TE$). }
\vspace{2mm}
\centering
\footnotesize
\begin{tabular}{|c|cc|}
\hline
Parameter & Model 1 & Model 2 \\
\hline
$w$ &  0.0060 & 0.0046 \\
$\log c_s^2$ & 0.58 & 0.22 \\
$h$  & 0.0025 & 0.0022 \\
$100\Omega_b h^2$ & 0.017 & 0.017 \\
$10\Omega_{cdm}$ & 0.016 & 0.014 \\
$\log 10^{10}\Delta^2_\mathcal{R}$ & 0.0047 & 0.0047 \\
$n_s$  & 0.0032 & 0.0032\\
\hline
\end{tabular}
\label{tab:Euclid+planck}
\end{table}

Importantly, while Planck alone has no constraining power whatsoever on the dark energy sound speed, because of the breaking of parameter degeneracies discussed above,
 the combination of Planck and Euclid turns out to improve the Euclid-alone sensitivity to $\log c_s^2$ by about a factor of two.  Parameters such as the physical baryon density $\Omega_b h^2$ and the scalar spectral index $n_s$, to which Planck is most sensitive, also benefit somewhat from a combined analysis with Euclid: the sensitivity to $n_s$ from the combined improves on the Planck-alone result by about a factor two, while for $\Omega_b h^2$ we find a $\sim 30\%$ improvement.

\section{Discussions and conclusions\label{sec:V}}

Current observation of an accelerated universal expansion is commonly attributed to the presence of a dark energy component in universe's total
energy budget.  However, apart from that it should have some negative pressure,
our understanding of the details of this dark energy component is  quite limited. In this work, we have investigated how future cluster surveys
can help to shed light on this dark energy through their potential to observe galaxy clusters numbering in the hundred thousands.
We have considered in particular how  the cluster survey of the ESA Euclid project
will allow us to determine the equation of state and the sound speed of a generic dark energy fluid.

The main role of a dark energy sound speed is that it enables the clustering of dark energy on length scales above the sound horizon, thereby allowing
the dark energy component to participate actively in the formation of structure in the universe.
This means that, if the sound horizon is comparable to or smaller than the typical length scales of a galaxy or galaxy cluster,
one should expect to find dark energy bound gravitationally in the collapsed objects and contribute to their mass.
We have demonstrated in this work, using the spherical collapse model, that the maximum dark energy contribution to the total cluster mass is of order a few tenths of a per cent at the time of virialisation.

Secondly, because dark energy clustering feeds back on the evolution of the dark matter density perturbations, it leads to a scale-dependence in the dark matter clustering that turns out to be more significant than is suggested by the mere $\sim 0.1\%$  infall discussed immediately above.    For dark energy sound speeds approximately in  the range $c_s^2 =10^{-6} \to 10^{-4}$,
the effect of dark energy clustering is directly visible in the cluster mass function in the mass range $10^{12} \to 10^{16} M_\odot$; for dark energy models consistent with present observations,  the number massive  clusters  ($\sim 10^{16} M_\odot$) can potentially change by as much as $15\%$ when compared with the no clustering case.

Using our predictions for the cluster mass functions, we have performed a Fisher matrix forecast in order to assess the potential of the Euclid cluster survey to constrain dark energy parameters.
We find that a $1 \sigma$ sensitivity to the (time-independent) dark energy energy equation of state parameter $w$ at the sub-percent level is possible, especially when the
 the Euclid cluster survey is complemented by CMB observations from Planck.  Furthermore,  if the dark energy has a sound speed that deviates appreciably from unity, we will be able to
 pin it down with future cluster surveys to roughly within an order of magnitude.

The results presented in this work are well in line with those from previous studies, where observational probes other than cluster surveys had been considered~\cite{Hu:2004yd,Takada:2006xs,Xia:2007km,Abramo:2009ne,Ballesteros:2010ks}.
We therefore conclude that future cluster surveys will be competitive with other probes in their constraining power on dark energy parameters,
and that the optimal strategy is a combination of different observations.

\ack

OEB thanks the Villum Foundation for support. TB thanks Thomas Tram for assistance with CLASS.

\section*{References}

\bibliographystyle{unsrt}
\bibliography{refs}

\end{document}